\documentclass[useAMS,usenatbib]{mn2e}

\usepackage{graphicx}\usepackage{epsfig}\usepackage{epsf}
\usepackage{amsmath}\usepackage{amssymb}\usepackage{stfloats}
\title[SN~2009ip]{Massive stars dying alone: The extremely remote
  environment of SN~2009ip}

\author[Smith]{Nathan Smith$^{1}$\thanks{E-mail:
    nathans@as.arizona.edu}, Jennifer E. Andrews$^1$, and Jon C.\
  Mauerhan$^2$ \\
  $^{1}$Steward Observatory, University of Arizona, 933 N. Cherry
  Ave., Tucson, AZ 85721, USA \\
  $^2$Department of Astronomy, University of California, Berkeley, CA
  94720-3411, USA}

\begin{document}

\pagerange{\pageref{firstpage}--\pageref{lastpage}} \pubyear{2012}
\maketitle
\label{firstpage}

\begin{abstract}

  We present late-time {\it Hubble Space Telescope} ({\it HST}) images
  of the site of supernova (SN) 2009ip taken almost 3 yr after its
  bright 2012 luminosity peak.  SN~2009ip is now slightly fainter in
  broad filters than the progenitor candidate detected by {\it HST} in
  1999.  The current source continues to be dominated by ongoing
  late-time CSM interaction that produces strong H$\alpha$ emission
  and a weak pseudo-continuum, as found previously for 1-2 yr after
  explosion.  The intent of these observations was to search for
  evidence of recent star formation in the local ($\sim$1kpc;
  10{\arcsec}) environment around SN~2009ip, in the remote outskirts
  of its host spiral galaxy NGC~7259.  We can rule out the presence of
  any massive star-forming complexes like 30 Dor or the Carina Nebula
  at the SN site or within a few kpc.  If the progenitor of SN~2009ip
  was really a 50-80 $M_{\odot}$ star as archival {\it HST} images
  suggested, then it is strange that there is no sign of this type of
  massive star formation anywhere in the vicinity.  A possible
  explanation is that the progenitor was the product of a merger or
  binary mass transfer, rejuvenated after a lifetime that was much
  longer than 4-5 Myr, allowing its natal H~{\sc ii} region to have
  faded.  A smaller region like the Orion Nebula would be an
  unresolved but easily detected point source.  This is ruled out
  within $\sim$1.5 kpc around SN~2009ip, but a small H~{\sc ii} region
  could be hiding in the glare of SN~2009ip itself.  Later images
  after a few more years have passed are needed to confirm that the
  progenitor candidate is truly gone and to test for the presence of a
  small H~{\sc ii} region or cluster at the SN position.

\end{abstract}

\begin{keywords}
  circumstellar matter --- stars: evolution --- stars: winds, outflows
  --- supernovae: general --- supernovae: individual (2009ip)
\end{keywords}

\begin{figure*}
\includegraphics[width=5.8in]{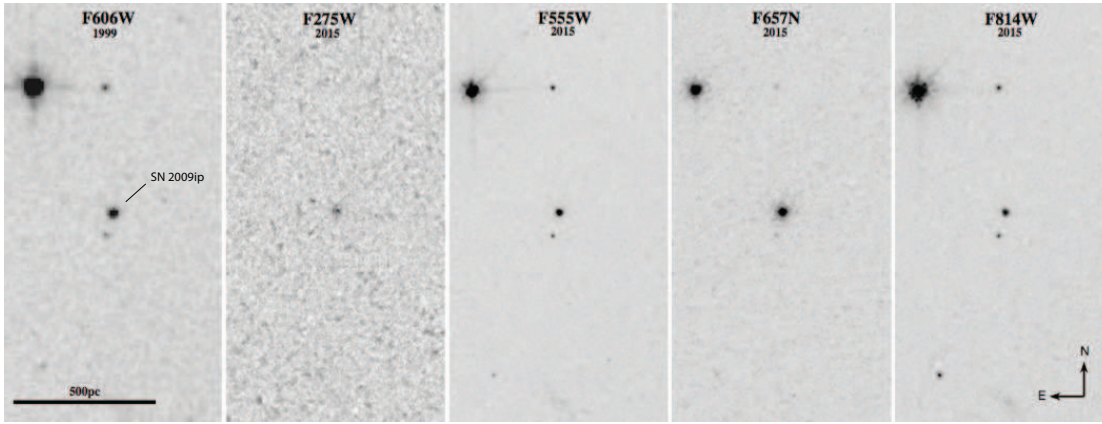}
\caption{{\it HST} images of the immediate environment of SN~2009ip.
  The left panel is the same WFPC2 F606W image of the progenitor
  obtained in 1999, from \citet{smith10}.  The other four panels are
  the new WFC3-UVIS images in F275W, F555W, F657N, and F814W obtained
  in May 2015 (see Table~\ref{tab:phot}). SN~2009ip is the brighter of
  the two sources at the center of the image.}
\label{fig:img}
\end{figure*}

\section{INTRODUCTION}

The class of Type~IIn supernovae (SNe~IIn hereafter), whose narrow H
lines indicate strong interaction with dense circumstellar material
(CSM), have challenged our understanding of stellar evolution and
death.  Their dense H-rich CSM, progenitor instability, and high
initial masses inferred from various clues suggest a link to the class
of luminous blue variables (LBVs), which are not supposed to be
anywhere near core collapse in the standard scenario of massive
single-star evolution (see \citealt{smith14} for a general review).

Among well-studied examples of SNe~IIn, the explosion of SN~2009ip in
mid-2012 (note that its discovery in 2009 was deemed to be a SN
impostor) is one of the most interesting, with by far the best
observational characterization of a directly detected progenitor among
any SN in history (even SN~1987A).  It had a (presumably) quiescent
progenitor star candidate detected in archival {\it Hubble Space
  Telescope} ({\it HST}) images, with a luminosity that implied a very
high initial mass of at least 50-60 $M_{\odot}$
\citep{smith10,foley11}.  This source also showed a series of
outbursts in the few years before the SN \citep{smith10,pastorello13}
that were reminiscent of both S~Dor outbursts and giant eruptions of
LBVs.  Unlike any progenitor source so far, high-quality spectra of
these precursor outbursts were obtained, with a detailed analysis
before the SN indicating a strong similarity to LBVs
\citep{smith10,foley11}.\footnote{Lower-quality, low-resolution
  photographic spectra were identified for SN~1987A \citep{walborn89},
  but so far this is the only other SN with a progenitor spectrum.}
With a quiescent and very luminous progenitor, an S~Dor outburst,
several bright but brief SN impostor eruptions, and progenitor spectra
resembling LBVs, SN~2009ip provides a strong link between LBVs and
SNe~IIn.

The repeating variable source at the position of SN~2009ip began to
brighten again in mid-2012, but this time things were different.
Spectra of the fainter 2012a peak showed very broad P Cygni profiles
with velocities of 13,000 km s$^{-1}$, suggesting that the event was a
core-collapse SN and not another LBV outburst \citep{sm12,mauerhan13}.
The subsequent and brighter 2012b event showed a high peak luminosity
and a spectrum typical of SNe~IIn with strong CSM interaction.  The
2012 SN-like event has already been discussed extensively in the
literature \citep{mauerhan13,mauerhan14,pastorello13,prieto13,
  fraser13,fraser15,ofek13,raf14,smith13,smp14,graham14}.  In these
publications and in discourse at meetings, there was some uncertainty
and controversy about whether the 2012 event was a true core-collapse
SN, since (1) CSM interaction can provide bright transients even from
relatively low-energy explosions, (2) the initial SN was somewhat
fainter than standard SNe~II-P, and (3) the rich observational dataset
for the progenitor presented mysteries that were not easily explained
by any existing model.  These are, however, expressions of the
challenge in understanding SNe~IIn and CSM interaction, rather than
arguments against a core-collapse event.  While it is difficult to
prove definitively that the event was a core collapse because of the
masking of CSM interaction, a SN is the most straightforward
explanation of the data. \cite{smp14} showed that all available
evidence was consistent with the core collapse SN explosion of a blue
supergiant that encountered strong CSM interaction.  Moreover, both
line-profile evolution \citep{smp14} and spectrapolarimetry
\citep{mauerhan14} show that the CSM interaction was highly aspherical
and probably disk-like, forcing the kinetic energy budget of the event
to be $\sim$10$^{51}$ ergs.  \citet{emily14} also argued for a
disk-like CSM based on narrow line ratios.  \citet{fraser15} showed
that the source at +2 yr was consistent with steady ongoing CSM
interaction with no additional outbursts, adding further evidence in
favor of a core-collapse event.

SN~2009ip provides our clearest example of pre-SN instability that
leads to eruptive pre-SN mass loss in the few years before explosions,
which may be associated with the final nuclear burning sequences in
the last years of a massive star's life \citep{qs12,sa14}.
Alternative non-terminal models involving binary mergers and accretion
were also proposed for the 2012 event \citep{soker13,kashi13}, but
these cannot supply the required 10$^{51}$ ergs of kinetic energy.


In this paper, we are mainly concerned with the host galaxy
environment around SN~2009ip.  A fundamental interesting mystery was
that while progenitor detections pointed to a very massive unstable
star, the location of SN~2009ip was in the remote outskirts of its
spiral host, far away from obvious signs of recent star formation and
young stellar populations \citep{smith10,foley11,mauerhan13,raf14}.
It was located about 5 kpc from the center of its relatively small
host spiral galaxy NGC~7259.  By extrapolating the apparent
metallicity gradient measured in the inner $\sim$1.5 kpc out to the 5
kpc radius of SN~2009ip, \citet{raf14} infer a mildly subsolar
metallicity at the SN site of $0.4 < Z/Z_{\odot} < 0.9$.  SN~2009ip's
progenitor can therefore be compared with populations of massive stars
observed in the Milky Way and Large Magellanic Cloud (LMC).  An
interesting result is that despite their high luminosities and high
inferred initial masses, LBVs in the Milky Way and LMC appear to be
relatively isolated compared to expectations for their presumed role
in stellar evolution.  \citet{st15} demonstrated that LBVs selectively
avoid clusters of O-type stars, especially early O-types that are
their presumed progenitors.  More importantly, LBVs are more dispersed
on the sky than WR stars; this rules out the standard picture wherein
LBVs are a transitional phase between massive O-type stars and WR
stars.  Instead, \citet{st15} suggested that most LBVs may be the
result of interacting binary evolution, getting rejuvenated by either
mass transfer or mergers.  This would make them stand out as
anomolously young compared to their surrounding populations.  In other
words, {\it they are evolved massive blue stragglers}.  They may
become even more isolated upon receiving a kick from their companion
star's SN, although it is not yet clear if a kick is required to
explain their environments.  Also relevant to this story is that
SNe~IIn in general appear to be less correlated with bright H$\alpha$
in their host galaxies than other types of SNe
\citep{anderson12,habergham14}.  While this has been interpreted as
signifying lower initial masses by those authors, that interpretation
has been a topic of debate \citep{crowther13,st15}.  As described
below, SN~2009ip seems to follow a similar trend of not having bright
H$\alpha$ nearby, despite having a detection of a very massive and
luminous progenitor star.


\begin{table}
\begin{center}\begin{minipage}{3.3in}
    \caption{New {\it HST} WFC-UVIS Images of SN~2009ip, including ST
      magnitudes for SN~2009ip and background 3$\sigma$ upper
      limits}\scriptsize
    \begin{tabular}{@{}lccccc}\hline\hline
Date        &Filter &Exp.(s) &Mag  &1$\sigma$ &3$\sigma$ U.L. (mag) \\  \hline
2015 May 25 &F275W  &2900    &22.53 &0.05  &25.0  \\
2015 May 23 &F555W  &1650    &21.91 &0.01  &26.8  \\
2015 May 25 &F657N  &5911    &18.97 &0.003  &25.2  \\
2015 May 23 &F814W  &1100    &22.91 &0.01  &27.0  \\
\hline
\end{tabular}\label{tab:phot}
\end{minipage}\end{center}
\end{table}

\begin{figure}
  \includegraphics[width=3.2in]{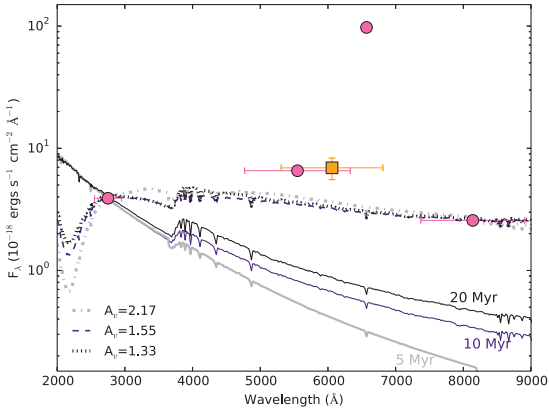}
  \caption{The photometry of the F606W progenitor candidate in 1999
    (orange square) and the photometry of the source detected in new
    {\it HST} images in 2015 (pink circles).  These are compared to
    Starburst99 models as discussed in the text for 5, 10, and 20 Myr
    (gray, blue, and black, respectively).  Models with these ages and
    no reddening (solid) are scaled to the F275W point, and the same
    models with reddening (dashed/dotted) are scaled to match the
    F275W and F814W points. See Table~\ref{tab:s99} and text.}
\label{fig:sed}
\end{figure}

\section{OBSERVATIONS}

The site of SN~2009ip was imaged with {\it HST} in 1999 using the
F606W filter on WFPC2.  A candidate progenitor source was detected
within the error circle of SN~2009ip.  The observations were presented
and analyzed by \citet{smith10} and \citet{foley11}.  From these data,
both investigations inferred the existence of a massive progenitor,
with an implied initial mass (compared to single-star evolutionary
tracks) of 50-80 $M_{\odot}$ or $>$60 $M_{\odot}$,
respectively.\footnote{As in \citet{smith10}, we adopt a distance
  modulus of $m-M=31.55$ mag, and a Galactic reddening and extinction
  of $E(B-V)$=0.019 mag and $A_R$=0.05 mag, respectively, for
  NGC~7259.}  

Independent of stellar evolution models, a minimum luminosity (i.e. no
bolometric correction or local extinction correction) of $>$10$^{5.9}$
$L_{\odot}$ \citep{smith10} would imply that the quiescent progenitor
would need to be at least 25 $M_{\odot}$ in 1999, assuming that it was
not exceeding the classical Eddington limit at that time.  It would
have had a significantly higher inital mass due to stellar wind mass
loss during its life.  The fact that it suffered an S~Dor-like
eruption after this -- followed by more extreme SN impostor-like
luminosity spikes in 2009-2011 --- lend weight to the idea that the
{\it HST} progenitor in 1999 may have been the quiescent star,
although the implied mass remains uncertain.

We obtained new images of the same location at almost 3 yr after the
2012 SN event using several filters with the WFC3-UVIS camera.
Observational parameters are summarized in Table~\ref{tab:phot}, which
includes photometry for SN~2009ip itself, as well as 3$\sigma$ upper
limits estimated from the background noise in the images in the region
around SN~2009ip.  We used three broad-band filters to sample the
UV/optical continuum (F275W, F555W, and F814W), plus F657N to sample
low-velocity H$\alpha$ emission at the host redshift.  The photometry
corresponds to days 1033 and 1035 after the date of discovery of the
2012a burst, following \citet{mauerhan13}.  The data for each filter
were corrected for charge transfer efficiency (CTE) trails using the
software tools available through STScI\footnote{{\tt
    http://www.stsci.edu/hst/wfc3/tools/cte\_tools}.}  Cosmic rays were
removed using the LA-COSMIC Package \citep{vand01} using suggested
paramaters.  The final CTE and CR corrected images were combined and
corrected for distortion effects using AstroDrizzle\footnote{See {\tt
    http://www.stsci.edu/hst/HST\_overview/drizzlepac}.}

The extremely remote environment shown in the new {\it HST} images is
surprising, as discussed below.  To help interpret the lack of any
diffuse H$\alpha$ emission detected in the vicinity of SN~2009ip, we
injected fake sources into the F657N image to illustrate how bright
various types of H~{\sc ii} regions should be.  We obtained flux
calibrated images of 30 Dor and the Carina Nebula from the Southern
H-Alpha Sky Survey Atlas (SHASSA) \citep{shassa} and scaled these as
they would appear if they were located in the outer parts of NGC~7259
at a distance of 20.4 Mpc.  The SHASSA image of the Orion Nebula is
badly saturated in the Huygens region around the Trapezium, so we used
an image of the Orion Nebula obtained with the WFI camera at the
ESO/MPIA 2.2m telescope on the nights of Jan 1-2, 2005
\citep{dario09}, which was made available through the {\it HST}
archive as a product from the Treasury project on
Orion\footnote{https://archive.stsci.edu/prepds/orion/}.

We also obtained a late-time spectrum of the site of SN~2009ip using
the Bluechannel spectrograph on the MMT on 2015 Nov 2, which
corresponds to day 1196 after the date of discovery.  This spectrum
was obtained with the 1200 lpm grating covering the wavelength range
from Na I D to H$\alpha$, and was reduced using standard techniques.
This was the same as in our previous MMT spectra of SN~2009ip
\citep{mauerhan13,mauerhan14, smp14}.  The spectrum shows strong
intermediate-width ($\pm$1000 km s$^{-1}$) H$\alpha$ emission, but any
continuum or other emission features are below the noise level. The
purpose of this spectrum was mainly to estimate the H$\alpha$ line
strength at late times so that we could appropriately scale a deeper
spectrum of an old SN~IIn at a similar time after explosion (in this
case we chose SN~2005ip; see below).

\begin{table}
\begin{center}\begin{minipage}{3.3in}
    \caption{Starburst99 clusters matched to SN~2009ip photometry (see
      Figure~\ref{fig:sed}), with TAMS turnoff mass ($M_{\rm TAMS}$),
      absolute $V$ magnitude of the cluster ($M_V$), and total cluster
      mass ($M_{cl}$).  $E(B-V)$ is the amount of reddening for the
      reddened models in Figure~\ref{fig:sed}, and the last two
      columns give the corrected $M_V$ and mass of the reddened
      cluster.}  \scriptsize
\begin{tabular}{@{}lcccccc}\hline\hline
 Age &$M_{\rm TAMS}$ &$M_{V}$ &$M_{cl}$ &$E(B-V)$ &$M_{Vcorr}$ &$M_{corr}$ \\
 (\it{Myr}) &($M_{\odot}$) &(mag) &($M_{\odot}$) &(mag) &(mag)  &($M_{\odot}$) \\
 \hline
    5  &54 &-6.99 &15  &0.70 &-9.14 &800\\
    10 &21 &-7.42 &40  &0.50 &-9.02 &710\\
    20 &12 &-7.65 &100 &0.43 &-8.97 &1120\\
   \hline
\end{tabular}\label{tab:s99}
\end{minipage}\end{center}
\end{table}

\begin{figure}
\includegraphics[width=3.2in]{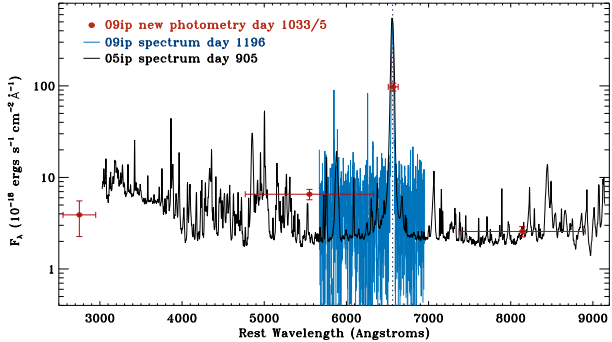}
\caption{The new {\it HST} photometry from Table~\ref{tab:phot}
  compared to a roughly contemporaneous spectrum of SN~2009ip (blue)
  taken with the MMT in 2015 November, as well as a late-time day 905
  spectrum of SN~2005ip (black) from \citet{smith09}.  SN~2005ip had
  stronger CSM interaction at almost 3 yr after explosion, and a
  spectrum with much higher signal to noise showing the fainter
  emission features that are expected to accompany strong H$\alpha$
  emission.  This SN~2005ip spectrum has been scaled to match the same
  H$\alpha$ flux as SN~2009ip.}
\label{fig:spec}
\end{figure}

\begin{figure}
\includegraphics[width=3.2in]{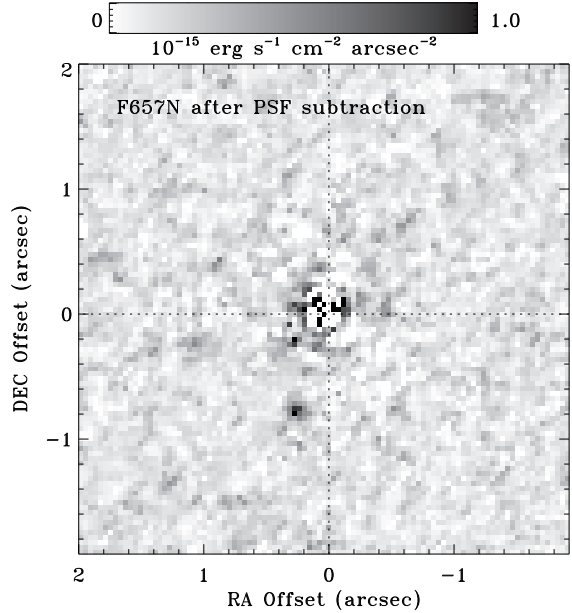}
\caption{The new F657N image of the location of SN~2009ip after
  subtraction of a model PSF made with TinyTIM.  There is residual
  extended emission at a level of roughly 0.5$\times$10$^{-15}$ ergs
  s$^{-1}$ cm$^{-2}$ arcsec$^{-2}$, but there are negative residuals
  at a similar level, so these could be subtraction artifacts. There
  is no extended emission beyond 0$\farcs$4 (40 pc) from the SN site
  brighter than $\sim$0.2$\times$10$^{-15}$ ergs s$^{-1}$ cm$^{-2}$
  arcsec$^{-2}$.}
\label{fig:psf}
\end{figure}

\section{RESULTS}

\subsection{SN~2009ip Late-time photometry}

The new {\it HST} images show a clearly detected point source at the
location of SN~2009ip. Based on the measured offset from a nearby
bright star in the field, this source is coincident with the position
of the 1999 progenitor to within 0.043 arcsec (roughly 10\% of a WFC3
pixel).  The late-time source has a profile consistent with the point
spread function in all the broad filters.  Figure~\ref{fig:psf} shows
the result of subtracting a model point spread function (PSF) from the
F657N image of SN~2009ip.  The model PSF was generated using the Tiny
Tim program\footnote{{\tt
    http://tinytim.stsci.edu/cgi-bin/tinytimweb.cgi}}.  There is
residual emission after the PSF subtraction at a typical brightness
level of 0.5$\times$10$^{-15}$ ergs s$^{-1}$ cm$^{-2}$ arcsec$^{-2}$
located within 0$\farcs$4 (40 pc) of the SN site.  This could indicate
some faint extended H~{\sc ii} region emission from a fading star
forming region.  However, there are negative residuals at a similar
level, so these are also consistent with being simple subtraction
artifacts due to an imperfect PSF match.  There is no detected
emission brighter than $\sim$2$\times$10$^{-16}$ ergs s$^{-1}$
cm$^{-2}$ arcsec$^{-2}$ beyond 40 pc from the SN site.  We also tried
using a PSF made from a bright star on the same image, with similar
results.  PSF subtraction of the broadband images (not shown) yielded
similarly inconclusive results.  In the continuum images, the
residuals allow for the presence of a faint SN light echo or extended
and very faint OB association that is 2-5\% as bright as the SN
point source, but as with F657N, the residuals are also consistent
with subtraction artifacts from an imperfect PSF match.

The F555W source is slightly fainter than the F606W progenitor, but
consistent with no change in the continuum level within the
uncertainty.  In a preliminary report of ground-based photometry,
\citet{thoene15} also reported that SN~2009ip continued to fade, and
by a few months after our {\it HST}images were taken, it was fainter
than the progenitor.

The faint emission detected in the broad filters could be dominated by
continuum from an underlying star cluster, late-time CSM interaction
emission from the fading SN itself, or a combination of these two.  In
fact, there must be some contribution from CSM interaction emission,
due to the strong and broad H$\alpha$ detected in the late-time
spectrum (Fig.~\ref{fig:spec}).  \citet{fraser15} showed that there
was still strong CSM interaction at around 2 yr after explosion.  We
can therefore rule out the possibility that there is no change in the
brightness of the progenitor candidate (as would be the case if it was
a chance alignment with an unrelated star), and so the progenitor must
have faded by whatever amount is now contributed by CSM interaction
luminosity.  The possibility that a progenitor survived but is now
obscured by dust is hard to rule out because it requires some
arbitrary amount of extinction to compensate for CSM interaction.
However, \citet{fraser15} noted that there was no clear indication of
new dust formation in the late-time data for SN~2009ip, so this would
argue that the progenitor candidate has in fact faded.

Figure~\ref{fig:sed} compares the observed spectral energy
distribution (SED) in the {\it HST} photometry to the hypothetical
emission from underlying young star clusters using examples from
Starburst99 models \citep{claus99}. We show models with ages of 5, 10,
and 20 Myr scaled to the F275W point only (assuming the the
optical/red emission is dominated by the SN), as well as these same
cluster model spectra with reddening to match both the F275W and the
F814W flux.  Clusters much older than 20 Myr would have turn off
masses too low (less than 12 $M_{\odot}$). The associated cluster
terminal-age main sequence masses ($M_{TAMS}$), absolute $V$ mags, and
total cluster stellar masses are listed in Table~\ref{tab:s99}.  The
corresponding absolute $V$ mag and total cluster mass corrected for
extinction are also shown, along with the corresponding $E(B-V)$
value.

If we assume that the CSM interaction dominates the optical/red
emission, the unreddened clusters allowed by the F275W photometry
imply total clutser masses that are unreasonably low (lower than the
progenitor mass of an individual star; this is because they have been
artificially scaled to match the photometry).  We discount these.
Adding reddening to the host cluster, we can account for the F275W and
F814W fluxes reasonably well, but there is a F555W excess flux of
about a factor of 2 above the model spectra, which could perhaps be
accounted for by SN emission (see below).  The stellar masses implied
are more reasonable, but the problem with these reddened clusters is
that the required reddening is larger than the value of the total
$E(B-V)$ inferred for the line of sight ($E(B-V)$=0.019 mag;
\citealt{smith10}).  A dominant contribution from a reddened cluster
therefore seems unlikely as well.  Overall, we find that an underlying
star cluster does not provide a compelling explanation for the
observed photometry, but we cannot rule out some contribution from a
small cluster or association hidden by SN~2009ip.

Another possibility is that the current photometry of SN~2009ip is
dominated by late-time SN emission from ongoing CSM
interaction. Figure~\ref{fig:spec} shows the SN~2009ip photometry
compared to a spectrum that we obtained at the MMT several months
later in 2015 November.  This spectrum indicates that very strong
H$\alpha$ emission with a resolved width of about 2000 km s$^{-1}$
continued to be observed, requiring the presence of strong ongoing CSM
interaction at these late times.  For comparison,
Figure~\ref{fig:spec} shows a late time (day 905) spectrum of
SN~2005ip from \citet{smith09}.  SN~2005ip was a Type~IIn with very
strong late-time CSM interaction, and this has much higher signal to
noise across a wide wavelength range that demonstrates the lower-level
emission in many lines that may accompany bright H$\alpha$.  Many of
these fainter emission lines blend together to make a blue
``pseudo-continuum'' \citep{smith09}.  The SN~2005ip spectrum has been
scaled to match the same H$\alpha$ flux as measured in the spectrum of
SN~2009ip, providing a rough indication of the pseudo-continuum
emission we might expect from SN~2009ip.  We do not necessarily expect
the late-time spectrum of SN~2009ip to be exactly the same as that of
SN~2005ip.  The point of this comparison is to illustrate that the
level of CSM interaction we detect via the H$\alpha$ emission in
SN~2009ip should be accompanied by a host of other emission lines
across a wide wavelength range that should contribute to the
broad-band filters.  Using SN~2005ip as a guide, it seems possible
that CSM interaction alone may account for the broadband flux and
relatively blue color.  A deeper spectrum of the late-time emission
can confirm this hypothesis.

Because the late-time CSM interaction emission must make some
contribution to the observed F555W photometry in 2015, we conclude
that any underlying continuum source other than CSM interaction should
be fainter than the F606W progenitor seen in 1999.  It is likely that
we will need to wait several more years to confirm that this source
has actually faded enough to place strong constraints on the
progenitor.  The main goal of our proposed {\it HST} observations was
to investigate the environment around SN~2009ip, as discussed next.

\begin{figure*}
\includegraphics[width=5.8in]{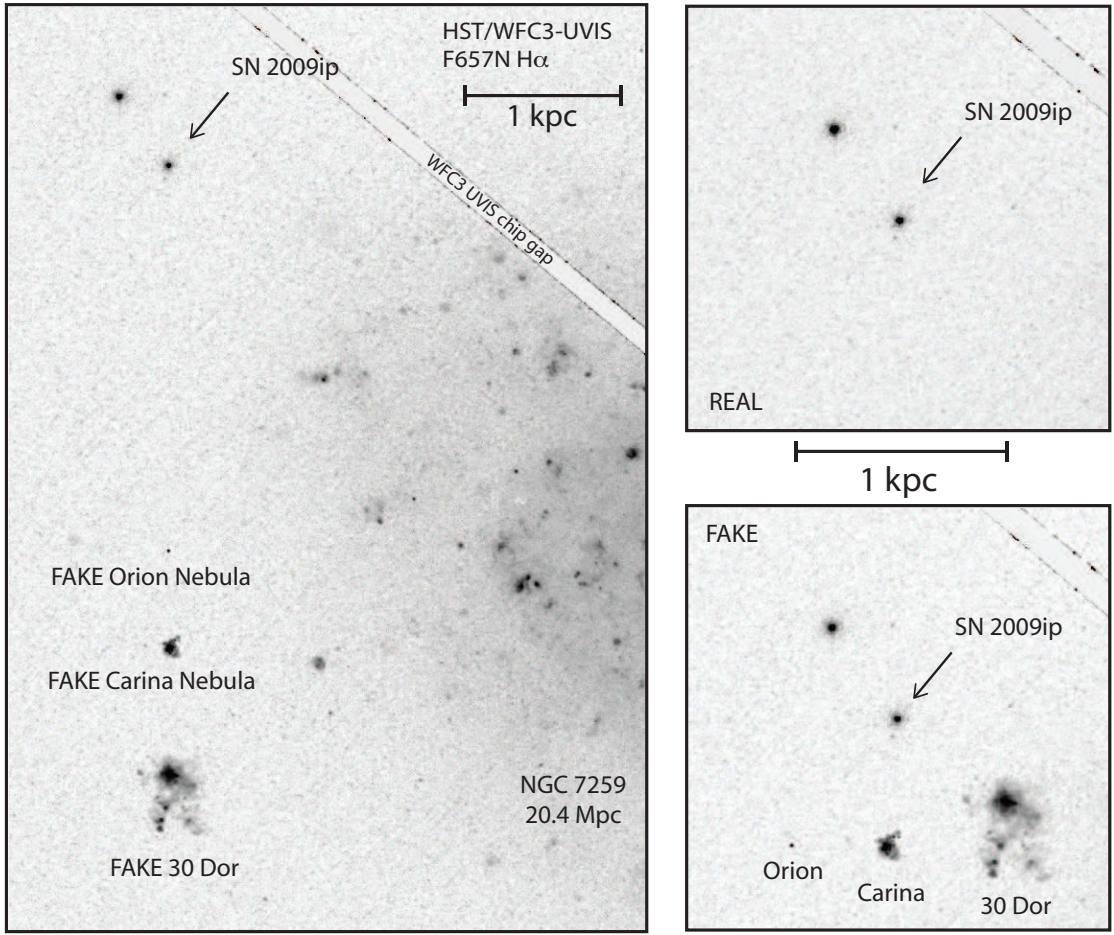}
\caption{HST/WFC3-UVIS images of SN~2009ip's environment in the F657N
  filter.  The left panel shows a wider view of the eastern half of
  the host galaxy, NGC~7259.  In the lower left corner, we have added
  in fake emission (made from real H$\alpha$ images scaled to the
  appropriate distance) demonstrating what the Orion Nebula, the
  Carina Nebula, and the 30 Doradus complex would look like if they
  were located in the outer portions of NGC~7259 and observed with the
  sensitivity of our image.  The two right panels show a zoomed in
  view of a 2 kpc box centered on SN~2009ip, with the real data (top)
  and a version with the same fake emission from Orion, Carina, and 30
  Dor added in for comparison.  The local environment of SN~2009ip
  shows essentially no evidence for recent massive star formation
  episodes in the form of young or aging H~{\sc ii} regions.}
\label{fig:carina}
\end{figure*}

\subsection{A Remote environment within 1 kpc}

The comparison between photometry of SN~2009ip and Starburst99 model
spectra in Figure~\ref{fig:sed} shows that our broad-band images could
easily detect any massive young clusters or OB associations in these
images.  The sensitivity determined from 3$\sigma$ upper limits in the
background of these images indicates that we would have detected
sources with $M_V$ of $-$4.7 mag, almost 100 times fainter than the
SN~2009ip point source.  This limit corresponds to a single mid to
late O-type main sequence star, or a single red supergiant.  The F275W
image rules out even relatively small clusters with ages younger than
10 Myr or any early O-type main sequence stars (initial masses above
40 $M_{\odot}$); these would be easily detected in the $\sim$1 kpc
environment around SN~2009ip.  No such bright clusters or luminous
stars are seen.  There is only one source about 100 pc to the south
east, which is, however, more than 10 times fainter than SN~2009ip at
most wavelengths, and not detected in F275W.  We take this as
indicating a remarkable lack of recent massive star formation in the
vicinity over the past $\sim$10 Myr.

H$\alpha$ also serves as a tracer of any recent or ongoing nearby star
formation.  Our F657N image was designed to be sensitive to even
relatively modest H~{\sc ii} regions.  Indeed, the F657N image reveals
dozens of H~{\sc ii} regions with a wide range of surface brightness
in the interior few kpc of the host galaxy nucleus, but there is a
marked absence of any H~{\sc ii} emission in the outer regions of the
galaxy beyond 3--4 kpc from the center.

The new F657N image of SN~2009ip and its surroundings is shown in
Figure~\ref{fig:carina}.  This image reveals no detectable H$\alpha$
emission from any young H~{\sc ii} regions in the surrounding $\sim$1
kpc around SN~2009ip, which is found about 5 kpc from the galaxy
center.  There is a bright point source to the north east (possibly a
foreground star) and two very faint point sources as well, but all
these other sources have similar brightness in broadband continuum
filters and so these are unlikely to be H~{\sc ii} regions.  SN~2009ip
itself is much brighter in the F657N filter, consistent with its
bright H$\alpha$ emission detected in the spectrum
(Fig.~\ref{fig:spec}).

To illustrate how the lack of H$\alpha$ detection indicates a dearth
of recent or ongoing massive star formation in the vicinity,
Figure~\ref{fig:carina} shows two panels that zoom-in on the
surrounding $\sim$1 kpc around SN~2009ip.  One of these is the
observed F657N image, and the other is the same image with some fake
comparison sources injected.  As noted earlier in Section 2, we used
H$\alpha$ images of the Orion Nebula, the Carina Nebula, and the 30
Dor complex, and scaled them to the appropriate pixel size as they
would appear in these images if they resided at the distance of
NGC~7259.  These sources were then convolved to the FWHM of the F657N
images and added to the FITS files of the {\it HST} data at the
appropriate surface brightness for the flux calibration in the {\it
  HST} images.  We also added these fake images to the left side of
the larger scale image in the same figure.  All three nebulae would
have been easily detected in the images, although the Orion Nebula is
so small that it would have been an unresolved point source (note
however, that the Orion Nebula is in a larger complex of star
formation that we did not include this in the fake injected image).

It is evident that the eastern side of NGC~7259 has no extreme
mini-starbursts comparable to 30 Dor, although there are a few
detected H$\alpha$ sources that have a scale in between 30 Dor and
Carina, and several comparable to Carina in the inner spiral arms of
the galaxy. There are a few dozen smaller H~{\sc ii} regions
comparable to Orion scattered around NGC~7259, but none in the
vicinity of SN~2009ip.

With a presumed initial mass of 50-80 $M_{\odot}$
\citep{smith10,foley11}, and a lifetime of only $\sim$4 Myr, one would
expect SN~2009ip to reside in a massive young cluster with many other
O-type stars, and to be surrounded by bright extended H~{\sc ii}
region emission.  This is clearly not the case, nor are there any such
H~{\sc ii} regons in the vicinity.  We dod not attempt to obtain deep
radio data to search for free-free emission from H~{\sc ii} regions or
SN remnants, although this may be interesting future observation to
make. We discuss implications corresponding to each type of region
below.

{\it 30 Dor:} \, Big H~{\sc ii} region complexes like 30 Dor can have
ongoing massive star formation for 10-20 Myr \citep{crowther13}.  30
Dor contains the very young (1--2 Myr) cluster R136, as well as the
$\sim$10 Myr old population that gave rise to SN~1987A.  If SN~2009ip
were in such a region, the new {\it HST} images would clearly detect
bright extended H$\alpha$ emission in the few 10$^2$ pc surrounding
SN~2009ip.  We can rule out the presence of such a large complex at
the position of SN~2009ip or within several kpc.

{\it Carina:} The carina nebula has birthed about 70 O-type stars
\citep{smith06}, and is of sufficient scale to sample the upper part
of the main sequence.  There are a handful of extremely massive stars
in Carina, including $\eta$ Carinae and three very massive WNH
stars. Moderately large H~{\sc ii} regions like Carina fade in 5 Myr
(Carina is now $\sim$3 Myr old) when the most massive O stars have
died and the bulk of the ionizing photons shut off.  With a presumed
initial mass of 50-80 $M_{\odot}$, one would expect SN~2009ip to come
from a region that is no older than than this (if it is a single
star).  However, we can clearly rule out a bright extended H$\alpha$
nebula like Carina at the position of SN~2009ip.  After about 5 Myr,
there may be a larger and fainter H$\alpha$ bubble.  This is harder to
rule out, and depends on an adopted escape fraction of ionizing
photons. As noted earlier, our PSF subtraction residuals allow for
some faint emission within 40 pc of the SN site, but these could also
be artifacts.  The lack of bright extended emission to a surface
brightness about 10 times fainter than Carina's outer shells
\citep{sb07} would imply a region significantly older than 5 or 6 Myr,
and would also therefore cause us to expect that any stars with
initial masses above 40 $M_{\odot}$ are long since dead.  The lack of
any Carina-like H~{\sc ii} region in the 1 kpc surrounding SN~2009ip
also suggests that it is unlikely to be a runaway from such a region.

{\it Orion:} The Orion Nebula contains only one O-type star (with an
initial mass around 30 $M_{\odot}$).  Smaller H~{\sc ii} regions like
this with only a couple O stars die after 5-6 Myr as well, when any
mid-O-type stars evolve off the main sequence and most of the ionizing
photons shut off.  The surrounding gas shells become neutral and get
much fainter and diffuse as they age.  Moreover, even when young and
bright, such a region would be very small in terms of angular extent
and would be unresolved by {\it HST} at 20 Mpc.  Such an H~{\sc ii}
region could easily be the birth environment of SN~2009ip and may be
hiding under the SN emission.  It would be unusual (although perhaps
not impossible with stochastic sampling of the initial mass function;
\citealt{jen14}) to find such a massive 50-80 $M_{\odot}$ progenitor
being born in this type of region.  It seems more plausible that such
an H~{\sc ii} region might give rise to a single very massive star
like the progenitor of SN~2009ip through binary evolution (see below).
Although an Orion-like region could be hiding underneath the current
photometry of SN~2009ip, it is important to note that such regions are
rarely found in complete isolation.  Massive star formation tends to
propagate from one region of a molecular cloud complex to another
\citep{el77}, leading to chains and complexes of H~{\sc ii} region
emission in disk galaxies.  Orion is part of a larger complex of star
formation in the Orion region, but no sign of any other nearby H~{\sc
  ii} regions are seen in the images of SN~2009ip.  Therefore, we have
the additional unusual circumstance that SN~2009ip's birth environment
would need to be a singular instance of star formation, or the last
such episode of massive star formation in the vicinity.

All of these considerations point to ages that are longer than 5 Myr,
and more likely around 10 Myr or more, suggesting initial masses
around 20 $M_{\odot}$.  This is in apparent contradiction with the
high luminosity of the progenitor star in its early phase before the
series of bright pre-SN eruptive events (at least 50-60 $M_{\odot}$).

One way to close the gap between the massive progenitor that was
detected and the surprising lack of recent massive star formation in
the vicinity would be to invoke binary evolution.  Either SN~2009ip
was a runaway star born at some other location, or it has been
rejuvenated by binary interaction.

If it is a runaway, the nearest indications of massive star formation
in NGC~7259 are found at least 1.5 kpc away.  To cover this distance
in a time of $\sim$3 Myr would require a kick speed of 400-500 km
s$^{-1}$.  This would be a very fast kick for a massive companion of a
SN.  Moreover, such high runaway speeds seem hard to reconcile with
the narrow emission lines in the spectrum of SN~2009ip (arising in the
pre-shock CSM), whose centroid radial velocities are consistent to
within $\la$100 km s$^{-1}$ with the redshift of the host galaxy
\citep{mauerhan13}.

A more likely alternative may be that the progenitor of SN~2009ip
increased its mass and luminosity late in life by accreting mass in a
mass transferring interacting binary system, or by a stellar merger
event (see
\citealt{pac71,dvb98,dvb13,podsiadlowski10,langer12,justham14}).  This
would allow the stars to live a longer core H-burning main sequence
lifetime of a lower-mass star, and to then get rejuvenated by mass
accretion or merging to produce the luminous progenitor source that
was detected in 1999.  This is the sort of evolutionary scenario
suggested for LBVs based on their remote environments \citep{st15},
although it should be noted that SN~2009ip's environment seems
unusually remote even among known LBVs.  The most isolated LBV known
in nearby galaxies is R71 in the LMC, which is 300 pc from any O-type
stars, but still on the outskirts of 30 Dor \citep{st15}.
\citet{st15} pointed out, however, that rejuvenation by mass accetion
or merging might produce an LBV-like star that appears very isolated
even without a kick if it is a blue straggler in a relatively small,
isolated, and aging star cluster.  This scenario requires the presence
of a relatively modest $\sim$10 Myr-old star cluster or association at
the position of SN~2009ip, which could be hidden by the current
ongoing CSM interaction.  Observations after the CSM interaction
luminosity fades (this may take several years) can eventually test
this hypothesis.

A binary scenario for the progenitor of SN~2009ip was invoked
previously, independent of the remote environment.  The disk-like CSM
inferred to exist around SN~2009ip \citep{mauerhan14,smp14,emily14}
would be consistent with the notion that some strong binary
interaction had occurred in the recent past.  The repeating brief
luminosity spikes in 2009-2011 we also attributed to binary close
encouters \citep{mauerhan13,mauerhan14,smith14}, either through direct
stellar collisions at periaston in an eccentric system as in the
$\eta$ Car system \citep{smith11} or potentially due to periastron
accretion events \citep{kashi13}.  The remote environment discussed
here strengthens the argument for a binary progenitor.  If the sort of
binary rejuvenation discussed above is appropriate for SN~2009ip, then
it is important to remember that the companion should still be there
if the binary system did not merge.  This companion might be combined
with any host cluster light in the late time photometry after the CSM
interaction luminosity fades.

If this sort of binary rejuvenation scenario is {\it not} the origin
of SN~2009ip, then its very luminous progenitor source is difficult to
understand.  To produce such a luminous, H-rich progenitor with a
strong wind and eruptive variability may require a more exotic
scenario invoving a binary system with a massive compact object
accreting from a companion, for example, which might then lead to a
core-collapse SN explosion.

A pulsational pair instability (PPI) eruption \citep{woosley07} is the
only proposed single-star mechanism to produce a non-terminal eruption
that could account for the total energetics of SN~2009ip's 2012 event,
which has been shown to require a kinetic energy well above 10$^{50}$
ergs because of the asymmetric CSM and the large ejecta mass implied
by the sustained broad-lined spectrum
\citep{mauerhan13,mauerhan14,smp14}.  However, typically the PPI will
have a large ejecta mass and expansion speeds $\la$5,000 km s$^{-1}$
\citep{woosley07}, slower than the observed 13,000 km s$^{-1}$ broad
lines in SN~2009ip's 2012a event \citep{mauerhan13}.  The most
energetic PPI events approaching 10$^{51}$ ergs would require
extremely high initial masses in excess of 100 $M_{\odot}$, and it is
uncertain if they occur at all at 0.5~$Z_{\odot}$ due to mass loss.
It seems very unlikely that there would be no evidence whatsoever of
recent massive star formation in the vicinity of SN~2009ip if it
originated from a star born with an initial mass above 100
$M_{\odot}$.

\smallskip\smallskip\smallskip\smallskip
\noindent {\bf ACKNOWLEDGMENTS}
\smallskip
\footnotesize

Support was provided by the National Aeronautics and Space
Administration (NASA) through HST grant GO-13787 from the Space
Telescope Science Institute, which is operated by AURA, Inc., under
NASA contract NAS5-26555.  Support was also provided by the National
Science Foundation (NSF) through grants AST-1210599 and AST-1312221 to
the University of Arizona.  This study used data products from the
Southern H-Alpha Sky Survey Atlas (SHASSA), which is supported by the
National Science Foundation.

\end{document}